# Magnetization reversal time of magnetic nanoparticles at very low damping

William T. Coffey,[1] Yuri P. Kalmykov,[2] and Serguey V. Titov[3]

[1]*Department of Electronic and Electrical Engineering, Trinity College, Dublin 2, Ireland*

[2]*Laboratoire de Mathématiques et de Physique (EA 4217), Université de Perpignan Via Domitia, F-66860, Perpignan, France*

[3]*Kotel'nikov Institute of Radio Engineering and Electronics of the Russian Academy of Sciences, Vvedenskii Square 1, Fryazino, Moscow Region, 141120, Russia*

**Abstract**

The magnetization reversal time of magnetic nanoparticles is investigated in the very low damping regime. The energy-controlled diffusion equation rooted in a generalization of the Kramers escape rate theory for point Brownian particles in a potential to the magnetic relaxation of a macrospin, yields the reversal time in closed integral form. The latter is calculated for a nanomagnet with uniaxial anisotropy with a uniform field applied at an angle to the easy axis and for a nanomagnet with biaxial anisotropy with the field along the easy axis. The results completely agree with those yielded by independent numerical and asymptotic methods.

# I. INTRODUCTION

A fine ferromagnetic particle is characterized by an internal potential, which has several local states of equilibrium with potential barriers between them. If the particles are small (~100 Å) so that the potential barriers are relatively low, the magnetization vector **M** may cross over the barriers between one potential well and another due to thermal agitation. The ensuing thermal instability of magnetization results in the phenomenon of superparamagnetism and in magnetic aftereffect.[1] The thermal fluctuations and relaxation of the magnetization of such nanoparticles play a central role in information storage, rock magnetism, magnetic hyperthermia, etc.[2] Furthermore, experimental success[3-6] in measuring the magnetization reversal time of an individual particle, and in verifying[3-6] its behavior as a function of the damping parameter predicted by the Néel-Brown theory[7-9] has stimulated renewed interest in the Kramers escape rate theory[10] as applied to classical macrospins. The Néel-Brown theory is in effect an adaptation of the Kramers theory[10,11] originally given for point Brownian particles to macrospin relaxation governed by a gyromagneticlike equation. Hence the verification of that theory in the magnetic nanoparticle[3-6] context nicely illustrates the Kramers conception of a thermal relaxation process over a potential barrier arising from the shuttling action of the Brownian motion.

Néel's original estimate[7] of the magnetization reversal time in a magnetic nanoparticle was based on transition state theory (TST)[11] yielding

$$\tau^{\mathrm{TST}} \sim f_A^{-1} e^{\Delta E}, \qquad (1)$$

where $\Delta E$ represents a dimensionless internal anisotropy barrier and $f_A$ is the so-called attempt frequency associated with the frequency of the gyromagnetic precession. Since $\Delta E$ is proportional to the volume of the particle, the relaxation time can vary from as little as $10^{-9}$ s to geological epochs; hence there is a fairly well-defined particle radius above which the magnetization will appear stable because the relaxation time greatly exceeds our own lifespan. However, Brown[8,9] criticized Néel's TST-based approach since gyromagnetic effects are not explicitly included and the damping dependence of the prefactor is ignored. The key to a more precise treatment of the reversal time lies in the construction of a Langevin equation for the evolution of the magnetization. Thus Brown proceeded by taking as a Langevin equation, Gilbert's equation for the motion of **M** augmented by a random field **h**(*t*) which may be written as[8,9]

$$\dot{\mathbf{u}}(t) = -\left[\mathbf{u}(t) \times \left[\gamma \mathbf{H}(t) + \gamma \mathbf{h}(t) - \alpha \dot{\mathbf{u}}(t)\right]\right]. \qquad (2)$$

Here $\mathbf{u} = \mathbf{M}/M_S$ is the unit vector directed along **M**, $M_S$ is the saturation magnetization, $\gamma$ is the gyromagnetic ratio, $\alpha$ is the dimensionless damping parameter, $\mathbf{H} = -(\mu_0 M_S)^{-1} \nabla V$ is the effective magnetic field comprising the anisotropy and external fields, the operator $\nabla = \partial/\partial \mathbf{u}$ indicates the gradient on the surface of the unit sphere, $V(\vartheta,\varphi)$ is the free energy density, the angles $\vartheta$ and $\varphi$ specify the orientation of $\mathbf{M}$ in spherical polar coordinates, and $\mathbf{h}(t)$ is a random magnetic field with Gaussian white noise properties. For some particular cases, e.g., for uniaxial particles with anisotropy energy density $K$ and easy axis along the polar axis, $V$ depends on $\vartheta$ only: $V(\vartheta) = -K\cos^2\vartheta$. Equation (2) is known as the *magnetic Langevin equation*. The random field accounts for the thermal fluctuations of the magnetization of an *individual* nanoparticle. Since the random torque counteracts the damping torque it can, if the temperature is high enough, cause magnetization reversal. Brown then derived from Eq. (2) the appropriate Fokker-Planck equation for the distribution function $W(\vartheta,\varphi,t)$ of the orientations of the magnetization vector $\mathbf{M}$ on the surface of the unit sphere,[8,9] viz.,

$$\frac{\partial W}{\partial t} = \frac{1}{2\tau_N}\left(\nabla^2 W + \frac{v}{kT}\left\{\alpha^{-1}\left(\mathbf{u}\cdot[\nabla W \times \nabla V]\right) + \left(\nabla\cdot(W\nabla V)\right)\right\}\right), \qquad (3)$$

where $\tau_N = v\mu_0 M_S(\alpha^{-1} + \alpha)/(2\gamma kT)$ is the free diffusion time of the magnetization, $v$ is the volume of the particle, $k$ is Boltzmann's constant, and $T$ is the temperature. Here, the particle is assumed to be uniformly magnetized. Although such a coherent rotation or "macrospin" approximation cannot explain all aspects of the magnetization dynamics in nanomagnets, many qualitative features needed to explain experimental data are satisfactorily reproduced. A detailed discussion of the assumptions made in deriving the Fokker-Planck and Gilbert equations is given elsewhere.[8,9,12-14]

The reversal time $\tau$, which is the longest relaxation time of the magnetization, may be defined as the inverse of the smallest nonvanishing eigenvalue $\lambda_1$ of the Fokker-Planck operator in Eq. (3).[8,9] The reversal time may be estimated using three different approaches: (i) Brownian or Langevin dynamics simulations, see, e.g., Refs. 15 and 16; (ii) numerical solutions of the Fokker-Planck equation, Eq. (3), see, e.g., Refs. 13 and 17–20; and (iii) analytical solutions of Eqs. (2) and (3) such as those yielded by the mean first passage time (MFPT), escape-rate theory, etc.; see, e.g., Refs. 8, 9, 14, and 20–24. These complementary approaches allow one to evaluate $\tau$ over wide ranges of temperature, damping, etc. In particular, numerical methods and escape rate theory are very useful for the determination of $\tau$ at low and high potential barriers, respectively. However, they have considerable limitations: for example, escape rate theory cannot be used at low barriers, $\Delta E \leq 1$, while numerical methods encounter substantial computational difficulties[15] in the very low damping (VLD) range, $\alpha \ll 1$, where the dynamics are almost entirely determined by the pure gyromagnetic Larmor equation. The VLD damping range is in practice the most important because both experimental and theoretical estimates yield small values of $\alpha$ of the order of 0.001-0.1; see, e.g., Refs. 3, 5, 12, and 25. For VLD, the TST equation (1) can considerably underestimate the true value of that time.[14]

In his earliest calculations of the reversal time, Brown[8] confined himself to uniaxial particles subjected to a dc external magnetic field $\mathbf{H}_0$ applied *along* the easy axis of the magnetization, where

$$V(\vartheta) = -\frac{kT\sigma}{v}(\cos^2\vartheta + 2h\cos\vartheta). \tag{4}$$

Here $\sigma = vK/(kT)$ is the dimensionless barrier height parameter, $K$ is an anisotropy constant, and $h = \mu_0 M_S H_0/(2K)$ is the external field parameter. In this axially symmetric situation, since no dynamical coupling between the longitudinal and the transverse modes of motion exists, the longitudinal relaxation is governed by a *single* state variable, namely, the polar angle $\vartheta$ of $\mathbf{M}$. The second state variable, namely, the azimuthal angle $\varphi$, gives rise only to a steady precession of $\mathbf{M}$. By recognizing this fact, Brown obtained from Eq. (3) a Fokker-Planck equation in $\vartheta$ only, viz.,[8]

$$\frac{\partial W}{\partial t} = \frac{1}{2\tau_N \sin\vartheta}\frac{\partial}{\partial \vartheta}\left[\sin\vartheta\left(\frac{\partial W}{\partial \vartheta} + \frac{vW}{kT}\frac{\partial V}{\partial \vartheta}\right)\right]. \tag{5}$$

We remark, that the *exact* Fokker-Planck equation (5) has the same mathematical form as the Debye *noninertial* rotational diffusion equation for polar molecules;[13] however, it is valid for *all values* of the damping parameter $\alpha$. This is so because Eq. (5) arises not from strong damping of the angular momentum as in the Debye diffusion equation rather it follows from the *axial symmetry* of $V(\vartheta)$. For *axially symmetric* potentials, Brown,[8] Aharoni,[17] and others (see Refs. 13 and 14 for a review) have developed various techniques such as variational methods,[8,11] MFPT,[11,20] etc. for the calculation of the reversal time. As an example, we mention Brown's well-known asymptotic formula for the reversal time, which becomes in the VLD limit[8]

$$\tau_{as}^{VLD} = \frac{\mu_0 M_S \sqrt{\pi/\sigma}e^{\sigma(1-h)^2}}{2\alpha\gamma K(1-h^2)\left[1-h+(1+h)e^{-4h\sigma}\right]}. \tag{6}$$

The reversal time can also be evaluated via the differential equation for the MFPT $\tau(\vartheta)$, viz.,[11,13,20]

$$L_{FP}^{\dagger}\tau(\vartheta) = -1 \tag{7}$$

with the appropriate boundary condition. Here $L_{FP}^{\dagger}$ is the adjoint Fokker–Planck operator associated with Eq. (5). The MFPT is the average time needed to reach the barrier point $C$ for the *first* time from a starting point $\vartheta$ inside the initial potential well (domain of attraction).[11] In particular, for the double-well potential with two nonequivalent wells $V(\vartheta)$, Eq. (4), the exact equations for the MFPTs $\tau_+ = \tau(\vartheta_A^+)$ and $\tau_- = \tau(\vartheta_A^-)$ from the minima of the deeper and shallow wells are given by[20]

$$\tau_\pm = \frac{\sigma\mu_0 M_S}{\alpha\gamma K} \int_{\vartheta_A^\pm}^{\vartheta_C} \frac{e^{\sigma(\cos^2\vartheta + 2h\cos\vartheta)}}{\sin\vartheta} \int_{\vartheta_A^\pm}^{\vartheta} e^{-\sigma(\cos^2\vartheta' + 2h\cos\vartheta')} \sin\vartheta' d\vartheta' d\vartheta. \quad (8)$$

Here $\vartheta_A^+ = 0$ and $\vartheta_A^- = \pi$ and $\vartheta_C = \arccos(-h)$ are the angular coordinates of the minima and maximum of $V(\vartheta)$, Eq. (4). Now recalling that $\tau_+$ and $\tau_-$ are related to the corresponding escape rates from the individual wells via[11] $\Gamma_+ = (2\tau_+)^{-1}$ and $\Gamma_- = (2\tau_-)^{-1}$ so that the overall reversal time is given by $\tau^{\text{VLD}} = (\Gamma_+ + \Gamma_-)^{-1}$, we have[13,14]

$$\tau^{\text{VLD}} = \frac{2\tau_+\tau_-}{\tau_+ + \tau_-}. \quad (9)$$

For high barriers $\sigma(1-h)^2 \gg 1$, $\tau^{\text{VLD}}$ from Eqs. (8) and (9) is closely approximated by the asymptotic equation (6).

Due to the mathematical difficulties encountered, the various methods developed for the axially symmetric case cannot be directly applied to the VLD reversal time of nanomagnets if the relaxation is governed by the general Fokker-Planck equation (3). These difficulties which arise because more than one space variable is now involved were overcome for the first time by Klik and Gunther.[23] In the high barrier limit, $\Delta E \gg 1$, they derived, via the uniform asymptotic expansion of the MFPT method of Matkowski *et al.*,[26] the VLD reversal time from an individual well with a *single* escape path for *non-axially symmetric free energy densities* $V(\vartheta,\varphi)$, viz.,[23]

$$\tau_{\text{as}}^{\text{VLD}} = \frac{\tau^{\text{TST}}}{\alpha S_{E_C}}, \quad (10)$$

where $\tau^{\text{TST}}$ is the TST reversal time, Eq. (1), and $S_{E_C}$ is the dimensionless action given by[23]

$$S_{E_C} = \frac{v}{kT} \oint_{E=E_C} \left( \frac{1}{\sin\vartheta} \frac{\partial V}{\partial\varphi} d\vartheta - \sin\vartheta \frac{\partial V}{\partial\vartheta} d\varphi \right), \quad (11)$$

The contour integral in Eq. (11) is taken along the critical energy trajectory, or separatrix, on which the magnetization may reverse by passing through the saddle point(s) of the energy $E_C$. The critical energy is the energy required by a spin to just escape the well and the separatrix delineates the bounded precessional motion in the well from that outside it. In the VLD regime, the system is only *very lightly* coupled to the bath so that the energy loss per cycle of the almost-periodic noisy motion of the magnetization on the saddle-point energy (escape) trajectory is much less than the thermal energy, $\alpha S_{E_C} \ll 1$, so that $\tau_{\text{as}}^{\text{VLD}} \gg \tau^{\text{TST}}$ for VLD. Everywhere the tacit assumption is made that the separatrix lies infinitesimally near to the *closed* noiseless and undamped trajectory with energy $E_C$.

Now the asymptotic equation (10) allows one to calculate the reversal time in nanomagnets with *nonaxially symmetric potentials* in the VLD regime, $\alpha S_{E_C} \ll 1$. However, Eq. (10) has the obvious drawback that it cannot be used for low barriers, $\Delta E \leq 1$. Moreover, the relation to the Kramers escape rate theory for point Brownian particles in the VLD range[10,11] is not immediately apparent. Nevertheless, both defects can be remedied via the energy-controlled diffusion equation for classical spins derived by Apalkov and Visscher[27] and Dunn *et al*.[28] Here we demonstrate how the VLD reversal time for spins can be evaluated from this equation via the MFPT[11,20] for *all* barrier heights and for *arbitrary* free energy density *V*. The MFPT method has been extensively applied to point particles and rigid rotators in Ref. 13, where the Hamiltonians are *separable* and *additive*. The generalization to the magnetic relaxation of macrospins with *nonseparable* and *nonadditive* Hamiltonians can now be accomplished because the energy-controlled diffusion equation for classical spins in the VLD limit is also a one-dimensional Fokker-Planck equation[27,28] like that for point particles. We remark that like point Brownian particles, in the escape rate problem as it pertains to spins, *three* regimes of damping appear. The latter arise as a direct consequence of the particular asymptotic method involved in the solution of the Fokker-Planck equation, namely, (i) intermediate-to-high damping (IHD) $\alpha S_{E_C} \gg 1$, VLD $\alpha S_{E_C} \ll 1$, and a more or less critically damped turnover range $\alpha S_{E_C} \sim 1$, where neither IHD nor VLD formulas apply.[10,11,14] In each range, the damping dependence of the escape rates, reversal time, etc. differ substantially. The interested reader can find a detailed discussion and appropriate formulas in Refs. 11, 13, and 14.

The present paper is arranged as follows. In Sec. II, we present the basic equations describing the stochastic spin dynamics in the VLD regime. In Sec. III, we derive in quadratures a general equation for the VLD reversal time for magnetic nanoparticles using the energy-controlled diffusion equation for spins in substantially the same manner as for point particles.[10,13] Here, we also demonstrate that in the high barrier approximation, $\Delta E \gg 1$, our exact integral result via the energy-controlled diffusion equation reduces to the asymptotic solution of Klik and Gunther,[23] Eq.(10), thus reconciling their solution with that from the Kramers theory. By way of illustration of our general results, which are valid for an arbitrary free energy, we determine in Secs. IV and V, the VLD reversal time of magnetic nanoparticles with uniaxial and biaxial anisotropies, respectively. In Secs. VI, we compare our analytical results both with independent numerical calculations and the asymptotes from escape rate theory. The Appendixes A, B, and C contain the details of the calculations.

## II. STOCHASTIC SPIN DYNAMICS IN THE VLD REGIME

By analogy with Kramers' derivation[10] of the energy-controlled diffusion equation for point particles in the VLD limit, one may parameterize[27,28] the instantaneous magnetization direction of a macrospin by the *slow* dimensionless energy variable $E = vV/(kT)$ and the *fast* precessional variable $\phi$ with period $2\pi$ running uniformly along a closed Stoner-Wohlfarth orbit of energy $E$.[28] In the VLD case, the energy varies very slowly compared to $\phi$. For the *slightly damped deterministic precession*, i.e., when the random field $\mathbf{h}(t) = 0$, the state variables $E$ and $\phi$ satisfy the equations of motion

$$\dot{E} = -\frac{v\mu_0}{kT}\left(\mathbf{H}\cdot\dot{\mathbf{M}}\right), \quad \dot{\phi} = \Omega_E, \tag{12}$$

where $\Omega_E = 2\pi f_E$ and $f_E$ is the frequency of precession in the potential well at a given energy $E$. We also denote the corresponding energy-dependent precession period as $P_E = 1/f_E$. This period can be calculated explicitly by taking a closed line integral along a Stoner-Wohlfarth orbit of constant energy $E$, viz.,[28]

$$P_E = \gamma^{-1} \oint_E \frac{\left(\left[\mathbf{H}\times\mathbf{M}\right]\cdot d\mathbf{M}\right)}{\left\|\left[\mathbf{H}\times\mathbf{M}\right]\right\|^2}.$$

Furthermore in order to treat the stochastic motion of the magnetization in the VLD limit, the Langevin equations for the variables $E$ and $\phi$ can be written as[28] (in our notation)

$$\dot{E} = F_1 + \left(\mathbf{g}_1\cdot\mathbf{h}\right), \quad \dot{\phi} = \Omega_E + \left(\mathbf{g}_2\cdot\mathbf{h}\right), \tag{13}$$

where $E$ and $\phi$ are now *random variables*,

$$F_1(E,\phi) = -\frac{v\mu_0\alpha}{\gamma M_S kT}\left(\dot{\mathbf{M}}\cdot\dot{\mathbf{M}}\right),$$

$$\mathbf{g}_1(E,\phi) = \frac{v\mu_0\Omega_E}{kT}\frac{\partial\mathbf{M}}{\partial\phi},$$

$$\mathbf{g}_2(E,\phi) = -\frac{v\mu_0\Omega_E}{kT}\frac{\partial\mathbf{M}}{\partial E}.$$

Equations (13), which are Langevin equations with multiplicative noise, describe the precession of the magnetization subject to *weak frictional* forces and *weak internal* fluctuations since $\alpha$ and $\mathbf{h}(t)$ are small. It follows that, in Eq. (13) and subsequently, $\dot{\mathbf{M}}$ must be understood in the *conservative* or purely Larmor sense as

$$\dot{\mathbf{M}}(t) = \gamma\left[\mathbf{H}\times\mathbf{M}\right]. \tag{14}$$

Dunn *et al.*[28] were then able to derive via the Langevin equations (13) interpreted in the Stratonovich sense[13] the Fokker–Planck equation for the probability density function $W(E,\phi,t)$. Since in the VLD regime, the energy $E$ diffuses very slowly over time, i.e., is almost conserved, while in contrast the phase $\phi$, which would be the only time-dependent variable in the completely conservative system, varies rapidly, the dependence on the fast variable $\phi$ may be eliminated. This is accomplished by averaging the probability density function $W(E,\phi,t)$ in energy-phase variables along a *closed* trajectory of the energy ultimately yielding the energy-controlled diffusion equation for the probability density function $W(E,t)$ in energy space,[27,28] viz.,

$$\frac{\partial W}{\partial t} = \alpha \frac{\partial}{\partial E}\left[S_E\left(f_E W + \frac{\partial}{\partial E}(f_E W)\right)\right], \quad (15)$$

where $S_E$ is the dimensionless action for spins given by

$$S_E = \frac{v\mu_0}{M_S kT}\oint_E ([\mathbf{H}\times\mathbf{M}]\cdot d\mathbf{M}) = \frac{v\mu_0}{\gamma M_S kT}\int_0^{1/f_E}|\dot{\mathbf{M}}|^2 dt. \quad (16)$$

We remark that the energy-controlled diffusion equation for spins, Eq. (15), is very similar but not identical to that for point Brownian particles in a potential $V(x)$.[10,11] The differences lie in the definitions of the damping coefficient and of the action. For point particles, they are $\alpha = \zeta/m$ and

$$S_E = \oint_E \sqrt{2m[kTE - V(x)]}dx = m\int_0^{1/f_E}|\dot{x}|^2 dt.$$

Here $E = [m\dot{x}^2/2 + V(x)]/(kT)$ is the normalized energy, $x$ and $m$ define the position and mass of a particle, respectively, $\zeta$ is the drag coefficient, and $f_E = kT\,\partial E/\partial S_E$ is the librational frequency. However, the calculation of $f_E$ and $S_E$ for spins is very much more involved than that for point particles because it must be carried out in spherical polar coordinates and the undamped motion is precession in space. Moreover, in the magnetization problem, the Stoner-Wohlfarth orbits,[28] namely, phase space orbits at constant energy inside the well, namely, have very complicated forms.[12]

## III. REVERSAL TIME IN THE VLD LIMIT

In order to evaluate the reversal time, we consider an assembly of spins in a potential well with a minimum at point *A*. In the true VLD case, $\alpha S_E \ll 1$, where the energy loss per cycle of a precessing spin is very much less than the thermal energy, the energy trajectories diffuse very slowly so that they do not differ significantly from those of the undamped precessional motion in a well. Then as a result of thermal fluctuations, on a noisy trajectory in the vicinity of the saddle energy the spin may have enough energy to escape over the potential barrier at the saddle point *C*. The energy-controlled diffusion equation for spins, Eq. (15), represents the continuity equation $\dot{W} + \partial_E J = 0$, where $J$ is the probability current. Now like the Kramers calculation[10] for particles (see also Hänggi et al.,[11] Section II.D), we consider the quasi-stationary solution of Eq. (15). Here with $\dot{W} = 0$ and $J(E) = J$ representing a steady injected current of spins to replenish those continually being lost at a saddle point $C$, we then find that the quasi-stationary distribution $W(E)$ satisfies the first-order linear differential equation

$$\frac{\partial}{\partial E}(f_E W) + f_E W = -\frac{J}{\alpha S_E}. \tag{17}$$

Next, considering the behavior of $W(E)$ at $E_C$ and following Kramers[10] and Hänggi et al.,[11] on assuming that $W(E_C) = 0$, i.e., all spins that reach the barrier go over, we have the particular solution of Eq. (17) as

$$W(E) = J \frac{e^{-E}}{\alpha f_E} \int_E^{E_C} \frac{e^{E'} dE'}{S_{E'}}. \tag{18}$$

In order to find the population $N$ in the well $A$, we integrate the quasi-stationary distribution $W(E)$ over the domain of the well energy so that

$$N = \int_{E_A}^{E_C} W(E) dE = \frac{J}{\alpha} \int_{E_A}^{E_C} \frac{e^{-E}}{f_E} \int_E^{E_C} \frac{e^{E'} dE'}{S_{E'}} dE$$

which becomes after integrating by parts

$$N = \frac{J}{\alpha} \int_{E_A}^{E_C} \frac{e^E}{S_E} \int_{E_A}^{E} \frac{e^{-E'} dE'}{f_{E'}} dE. \tag{19}$$

We then have via the flux-over-population method[11] the characteristic MFPT time $\tau^{\text{VLD}} = N/J$ from a potential well with energy $E_A$ over the saddle point $C$

$$\tau^{\text{VLD}} = \frac{1}{\alpha} \int_{E_A}^{E_C} \frac{e^E}{S_E} \int_{E_A}^{E} \frac{e^{-E'} dE'}{f_{E'}} dE. \tag{20}$$

This is the time to reach a separatrix from the point $A$ *provided that all spins there are absorbed*, which is the boundary condition that $W$ vanishes at $E = E_C$. The inverse of $\tau^{\text{VLD}}$ also determines the escape rate from the well.[11,26] Equation (20) can also be derived directly via the differential equation for the MFPT, viz.,[11,13,20] [c.f. Eq. (7)]

$$L_{\text{FP}}^\dagger \tau(E) = -1 \tag{21}$$

with the boundary condition $\tau(E_C) = 0$. The MFPT is the average time needed to reach the separatrix for the *first* time from a starting point $E_0$ inside the initial domain of attraction.[11] In the VLD limit, this time $\tau(E_0)$ becomes essentially independent of $E_0$, i.e., $\tau(E_0) \approx \tau^{\text{VLD}}$ for all starting configurations away from the neighborhood of the separatrix.[11] We emphasize that for the calculation of $\tau^{\text{VLD}}$ from Eq. (20) only a knowledge of the *deterministic dynamics* is required, i.e., $S_E$ and $f_E$ in Eq. (20) are always calculated via the deterministic Larmor equation (14), which can invariably be solved using either analytical or numerical methods.

The quadrature solution, Eq. (20), is valid for *all barrier heights* including *low barriers*, $\Delta E = E_C - E_A \leq 1$. However, in the high barrier limit, $\Delta E \gg 1$, Eq. (20) can be considerably simplified. Indeed, the main contribution to the inner integral of Eq. (20) comes from near the bottom of the well because the negative exponential dominates the integral in that region. Furthermore, the precession frequency now satisfies $f_{E_A} \approx f_A$, where $f_A$ is the well precession frequency which is independent of $E$ because of the paraboloid approximation for the potential near the bottom of the well [ $f_A$ is defined by Eq. (49) below]. Thus

$$\int_{E_A}^{E_C} \frac{e^{-E}}{f_E} dE \approx \frac{1}{f_A} \int_{E_A}^{\infty} e^{-E} dE = \frac{1}{f_A} e^{-E_A}. \quad (22)$$

In contrast, the main contribution to the outer integral of Eq. (20) comes from the positive exponential factor dominating the integrand near the saddle point $C$ of the potential. Therefore, noting Eq. (16) and using the approximation

$$S_{E'} \approx S_{E_C} = \frac{v}{kT} \oint_{E=E_C} \left( [\mathbf{u} \times \nabla V] \cdot d\mathbf{u} \right), \quad (23)$$

we have

$$\int_{E}^{E_C} \frac{e^{E'}}{S_{E'}} dE' \approx \frac{1}{S_{E_C}} \int_{-\infty}^{E_C} e^{E'} dE' = \frac{1}{S_{E_C}} e^{E_C}. \quad (24)$$

Using Eqs. (22), and (24) in Eq. (20) yields

$$\tau_{as}^{VLD} \sim \frac{e^{E_C - E_A}}{\alpha f_A S_{E_C}}. \quad (25)$$

In spherical polar coordinates $(\mathbf{e}_r, \mathbf{e}_\vartheta, \mathbf{e}_\varphi)$,[29] where $\mathbf{u} = \mathbf{e}_r$, $d\mathbf{u} = \mathbf{e}_\vartheta d\vartheta + \mathbf{e}_\varphi \sin\vartheta d\varphi$, and $\nabla V = \mathbf{e}_\vartheta \partial_\vartheta V + \mathbf{e}_\varphi \csc\vartheta \partial_\varphi V$, $S_{E_C}$ from Eq. (23) reduces to that given by Eq. (11). The contour integral in Eq. (23) is taken along the critical energy trajectory on which the magnetization may reverse by passing through the saddle point(s). In fact, Eq. (25) is just another equivalent representation of the TST equation (10) because $\tau^{TST} \sim (1/f_A) e^{E_C - E_A}$. Hence, our novel result, Eq. (20), predicts in the low-temperature limit exactly the same reversal time as Eq. (10) of Klik and Gunther.[23] In order to evaluate $\tau^{VLD}$ from Eq. (25), we require only explicit equations for $E_A$, $E_C$, $f_A$, and $S_{E_C}$. The calculation of the precession frequency $f_A$ and the well and saddle energies $E_A$ and $E_C$ is described in Appendix A, while the action $S_{E_C}$ can be calculated from Eq. (23) or Eq. (11). [13,14]

## IV. VLD REVERSAL TIMES FOR UNIAXIAL ANISOTROPY

As discussed above, Brown[8,9] calculated the reversal time for a uniaxial superparamagnet when a uniform magnetic field $\mathbf{H}_0$ is applied *along* the easy axis of the magnetization. However, by applying $\mathbf{H}_0$ at an angle $\psi$ with respect to the easy axis, the latter will also depend on the azimuthal angle $\varphi$ [cf. Eq. (4)]

$$E(\vartheta,\varphi) = -\sigma\left(\cos^2\vartheta + 2h\cos\psi\cos\vartheta + 2h\sin\psi\sin\vartheta\cos\varphi\right). \tag{26}$$

$E(\vartheta,\varphi)$ from Eq. (26) has a bistable structure with minima at $\mathbf{n}_A^+$ and $\mathbf{n}_A^-$ separated by a potential barrier with a saddle point at $\mathbf{n}_C$ (see Fig. 1). The saddle point lies generally in the equatorial region, while $\mathbf{n}_A^+$ and $\mathbf{n}_A^-$ lie in the north and south polar regions, respectively. In general, $E(\vartheta,\varphi)$ from Eq. (26) retains its *bistable* form for $0 \le h < h_c$, where $h_c = (\cos^{2/3}\psi + \sin^{2/3}\psi)^{-3/2}$ is some critical value of $h$ at which $E(\vartheta,\varphi)$ loses its bistable character.[30] When $\mathbf{H}_0$ is parallel to the easy axis, Eq. (26), the energy-landscape is a uniform equatorial ridge (zone) separating two polar minima and has no saddle point. A detailed treatment of the oblique-field problem has been given by Coffey *et al.*,[31] Kennedy,[32] Kalmykov *et al.*,[33,34] and Fukushima *et al.*[35] In particular, they showed that escape rate theory is in agreement with their numerical results, with computer simulations,[36] and with experiments,[5] emphasizing the vital importance of an accurate determination of the damping dependence of the reversal time. The non-axially symmetric double well potential, Eq. (26), is very instructive for the purpose of illustration of our principal result, Eq. (20), because accurate numerical results for the overall reversal time $\tau^{\text{VLD}}$ are already available for comparison in the literature.[33,34]

Now, to calculate the reversal time from the general equation (20), we need only the *deterministic* equations of motion of the magnetization. For $E(\vartheta,\varphi)$ from Eq. (26), the vector gyromagnetic equation (14) can be rewritten in terms of the Cartesian components $(u_X, u_Y, u_Z)$ of the unit vector $\mathbf{u}$ along the direction of magnetization $\mathbf{M}$ as

$$\tau_0 \dot{u}_X(t) = -[u_Z(t) + h\cos\psi] u_Y(t), \tag{27}$$

$$\tau_0 \dot{u}_Y(t) = [u_Z(t) + h\cos\psi] u_X(t) - h\sin\psi u_Z(t), \tag{28}$$

$$\tau_0 \dot{u}_Z(t) = h\sin\psi u_Y(t), \tag{29}$$

where

$$\tau_0 = \frac{\mu_0 M_S}{2\gamma K} \tag{30}$$

is a precession time constant. For $\gamma = 2.2 \cdot 10^5$ m A$^{-1}$s$^{-1}$, $\mu_0 = 4\pi \cdot 10^{-7}$ J m$^{-1}$ A$^{-2}$, $M_S \approx 1.4 \cdot 10^6$ A m$^{-1}$ and $K \approx 2 \cdot 10^5$ J m$^{-3}$ (cobalt), we have the estimate $\tau_0 \approx 2 \cdot 10^{-11}$ s. The solutions of Eqs. (27)-(29) are subject to the obvious constraint

$$u_X^2 + u_Y^2 + u_Z^2 = 1. \tag{31}$$

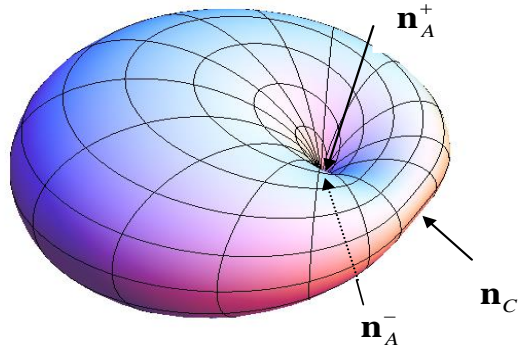

FIG. 1. 3D plot of $E(\vartheta,\varphi)/\sigma$, Eq. (26), for $h = 0.5$ and $\psi = \pi/2$.

Furthermore, the trajectories of the precessional dynamics must satisfy the additional constraint of energy conservation

$$\varepsilon = -u_Z^2 - 2h\cos\psi u_Z - 2h\sin\psi u_X = \text{const}, \tag{32}$$

where $\varepsilon = E/\sigma$ is the normalized free energy.

If $|h| < h_c \leq 1$, the potential from Eq. (26) has two *nonequivalent* wells with minima $\varepsilon_A^\pm$ and one saddle point at $\varepsilon_C$ (see Fig. 1). Both $\varepsilon_A^\pm$ and $\varepsilon_C$ can be presented as Taylor series[34] up to any desired order of $h$ [see Appendix A, Eqs. (53) and (54)]. Now, we must define two *individual* MFPTs, namely, $\tau_+$ from the deeper well ($\varepsilon_A^+ \leq \varepsilon \leq \varepsilon_C$) and $\tau_-$ from the shallow well ($\varepsilon_A^- \leq \varepsilon \leq \varepsilon_C$). These times are unequal in general, i.e., $\tau_+ \neq \tau_-$. Thus, Eqs. (16) and (20) as specialized to the non-axially symmetric double well potential from Eq. (26) yields for $\tau_+$ and $\tau_-$

$$\tau_\pm = \frac{\sigma^2}{\alpha} \int_{\varepsilon_A^\pm}^{\varepsilon_C} \frac{e^{\sigma\varepsilon}}{S_\varepsilon^\pm} \int_{\varepsilon_A^\pm}^{\varepsilon} \frac{e^{-\sigma\varepsilon'} d\varepsilon'}{f_{\varepsilon'}^\pm} d\varepsilon, \tag{33}$$

where

$$S_\varepsilon^\pm = 2\tau_0 \sigma \int_0^{1/f_\varepsilon^\pm} \left[ \dot{u}_{X\pm}^2(t) + \dot{u}_{Y\pm}^2(t) + \dot{u}_{Z\pm}^2(t) \right] dt. \tag{34}$$

Now, the anisotropy potential given by Eq. (26) has two nonequivalent wells so that this two-well nature of the potential must be accounted for[9,13,14] meaning that unlike in the isolated well *two* escape rates now exist. These comprise $\Gamma_+ = (2\tau_+)^{-1}$ from the deeper well and $\Gamma_- = (2\tau_-)^{-1}$ from the shallower well so that the overall reversal time $\tau^{VLD} = (\Gamma_+ + \Gamma_-)^{-1}$ is then given by Eq. (9).[13,14] Finally by solving the deterministic gyromagnetic equations (27)-(29) and then calculating the energy-dependent frequency $f_\varepsilon^\pm$ and dimensionless action $S_\varepsilon^\pm$ as described in Appendix B, $\tau^{\text{VLD}}$ can be determined from Eqs. (33) and (9). In the high barrier limit, $\sigma(\varepsilon_C - \varepsilon_A^-) \gg 1$, we have from Eqs. (9), (25), and (33),

$$\tau_{as}^{\text{VLD}} = \frac{2 e^{\sigma(\varepsilon_C - \varepsilon_A^-)}}{\alpha \left[ f_A^- S_{\varepsilon_C}^- + f_A^+ S_{\varepsilon_C}^+ e^{-\sigma(\varepsilon_A^- - \varepsilon_A^+)} \right]}. \tag{35}$$

Here the energy-independent precession frequencies $f_A^\pm$ in the vicinity of the *bottoms* of the wells and the dimensionless actions $S_{\varepsilon_C}^\pm$ pertaining to the *saddle* region are given by Eqs. (55) and (69) below, respectively.

For $\psi = 0$, i.e., if the external field $\mathbf{H}_0$ is applied *along* the easy axis of the magnetization so that the problem becomes axially symmetric, the above equations can be considerably simplified. By calculating $f_\varepsilon^\pm$ and $S_\varepsilon^\pm$ for $\psi = 0$ as described in Appendix B [Eqs. (70) and (71)], we then have from Eq. (33)

$$\begin{aligned}
\tau_\pm &= \frac{\tau_0 \sigma}{2\alpha} \int_{-1 \mp 2h}^{h^2} \frac{e^{\sigma \varepsilon} \int_{-1 \mp 2h}^{\varepsilon} \frac{e^{-\sigma \varepsilon'} d\varepsilon'}{\sqrt{h^2 - \varepsilon'}}}{\sqrt{h^2 - \varepsilon} \left( 1 + \varepsilon - 2h^2 \pm 2h\sqrt{h^2 - \varepsilon} \right)} d\varepsilon \\
&= \frac{\tau_0 \sqrt{\sigma \pi}}{2\alpha} \int_{-1 \mp 2h}^{h^2} \frac{e^{\sigma(\varepsilon - h^2)} \left\{ \text{erfi}\left[\sqrt{\sigma}(1 \pm h)\right] - \text{erfi}\left[\sqrt{\sigma(h^2 - \varepsilon)}\right] \right\}}{\sqrt{h^2 - \varepsilon} \left( 1 + \varepsilon - 2h^2 \pm 2h\sqrt{h^2 - \varepsilon} \right)} d\varepsilon.
\end{aligned} \tag{36}$$

where erfi($x$) is the error function of an imaginary argument defined as[37]

$$\text{erfi}(x) = \frac{1}{\sqrt{\pi}} \int_0^x e^{t^2} dt.$$

Notice that Eq. (36) reduces to Eq. (8) given above as it must. Furthermore, $\tau_{as}^{\text{VLD}}$ from Eq. (35) coincides with Eq. (6). In addition, for $h = 0$, via the transformation $\sigma\varepsilon \to -z^2$, Eqs. (9) and (36) yield $\tau^{\text{VLD}}$ for uniaxial nanomagnets in the absence of an external field, viz.,

$$\tau^{\text{VLD}} = \tau_0 \frac{\sigma \sqrt{\pi}}{\alpha} \int_0^{\sqrt{\sigma}} \frac{\text{erfi}(\sqrt{\sigma}) - \text{erfi}(z)}{\sigma - z^2} e^{-z^2} dz. \tag{37}$$

### V. VLD REVERSAL TIMES FOR BIAXIAL ANISOTROPY

By way of yet another practical illustration of Eq. (20), we consider a biaxial anisotropy potential augmented by the Zeeman term due to an external magnetic field $\mathbf{H}_0$ applied along the easy axis of magnetization, viz.,[21,22,38,39]

$$E(\vartheta, \varphi) = -\sigma \left( \cos^2 \vartheta - \delta \sin^2 \vartheta \cos^2 \varphi + 2h \cos \vartheta \right). \tag{38}$$

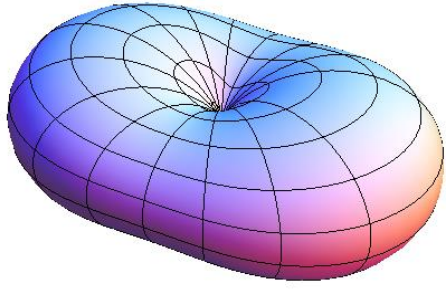

FIG. 2. 3D plot of $E(\vartheta,\varphi)/\sigma$, Eq. (38), for $\delta = 0.5$ and $h = 0.2$.

In general, $E(\vartheta,\varphi)$ from Eq. (38) has two *nonequivalent wells* and two *equivalent saddle points* (see Fig. 2). We remark that biaxial anisotropy may yield an appreciable contribution to the free energy density of magnetic nanoparticles.[6] In particular, Eq. (38) describes the free energy density of a spheroidal nanoparticle, with the axis of symmetry inclined at a certain angle to the easy anisotropy axis of the particle as well that of elongated particles, where easy- and hard-axis anisotropy terms are present.[22] Furthermore, the bistable potential in the form of Eq. (38) is commonly used in spintronic applications[12,27,40] in order to represent the free energy density of a nanopillar in the standard form of superimposed easy-plane and in-plane easy axis anisotropies. Finally, this example is very instructive for the purpose of illustration of our principal result, Eq. (20), because accurate numerical results for the reversal time for the biaxial potential are also available for comparison in the literature.[39]

For the biaxial anisotropy potential, Eq. (38), the gyromagnetic equation (14) can be written in terms of the Cartesian components $(u_X, u_Y, u_Z)$ of the unit vector $\mathbf{u}$ as

$$\tau_0 \dot{u}_X(t) = -[u_Z(t) + h] u_Y(t), \tag{39}$$

$$\tau_0 \dot{u}_Y(t) = [(1+\delta) u_Z(t) + h] u_X(t), \tag{40}$$

$$\tau_0 \dot{u}_Z(t) = -\delta u_X(t) u_Y(t). \tag{41}$$

The solutions of Eqs. (39)-(41) are again subject to the constraints of Eq. (31) and energy conservation

$$\varepsilon = -u_Z^2 - 2h u_Z + \delta u_X^2 = \text{const}, \tag{42}$$

where $\varepsilon = E/\sigma$ is the normalized free energy, the possible value of which is limited by the conditions $-1-2h \leq \varepsilon \leq \delta + h^2/(1+\delta)$.

If $|h| < 1$, the potential from Eq. (38) has two *nonequivalent* wells with minima $\varepsilon_A = -1 \pm 2h$ at $u_Z = \mp 1$ and two equivalent saddle points at $\varepsilon_C = h^2$ (see Fig. 2 and Appendix A). Thus, we must again define two individual MFPTs, namely, $\tau_+$ from the deeper well around the energy minimum at $u_Z = 1$ ($-1-2h \leq \varepsilon \leq h^2$) and $\tau_-$ from the shallow well around the energy minimum at $u_z = -1$ ($-1 + 2h \leq \varepsilon \leq h^2$). These times are again unequal in general, i.e., $\tau_+ \neq \tau_-$. Thus, Eq. (20), as specialized to biaxial anisotropy, becomes

$$\tau_\pm = \frac{\sigma^2}{\alpha} \int_{-1 \mp 2h}^{h^2} \frac{e^{\sigma\varepsilon}}{S_\varepsilon^\pm} \int_{-1 \mp 2h}^{\varepsilon} \frac{e^{-\sigma\varepsilon'} d\varepsilon'}{f_{\varepsilon'}^\pm} d\varepsilon, \qquad (43)$$

where $S_\varepsilon^\pm$ is defined by Eq. (34).

Now having solved Eqs. (39)-(41) and next having calculated $f_\varepsilon^\pm$ and $S_\varepsilon^\pm$ as described in Appendix C [see Eqs. (78) and (80)], the overall reversal time $\tau^{VLD}$ can be determined from Eqs. (9) and (43). Furthermore, in the high barrier limit, Eqs. (9) and (43) can once more be considerably simplified yielding Eq. (35), where $f_A^\pm$ and actions $S_{\varepsilon_C}^\pm$ are given by Eqs. (58) and (81), respectively. As described in Appendix C, Eq. (43) in the limit $\delta \to 0$ reduces to Eq. (36).

For $h = 0$, the biaxial free energy is a double-well potential with two *equivalent* wells, where only the region $-1 \leq \varepsilon \leq 0$ is appropriate because the energy of a separatrix trajectory is now $\varepsilon_C = 0$. Therefore, $\tau_+ = \tau_- = \tau$, so that the overall reversal time is then $\tau^{VLD} = \tau$. Having calculated $f_\varepsilon$ and $S_\varepsilon$ as described in Appendix C [Eqs. (82) and (83)], $\tau^{VLD}$ from Eq. (9) can be written as the analytic equation

$$\tau^{VLD} = \frac{\tau_0 \sigma}{2\alpha} \int_{-1}^{0} \frac{e^{\sigma\varepsilon} \int_{-1}^{\varepsilon} \frac{K(m_{\varepsilon'}) e^{-\sigma\varepsilon'}}{\sqrt{\delta - \varepsilon'}} d\varepsilon'}{\sqrt{\delta - \varepsilon} \left[ E(m_\varepsilon) + \varepsilon K(m_\varepsilon) \right]} d\varepsilon. \qquad (44)$$

Here $m_\varepsilon = \delta(1+\varepsilon)/(\delta-\varepsilon)$ and $K(m)$ and $E(m)$ are the complete elliptic integrals of the first and second kind, respectively.[37] For high barriers, $\sigma \gg 1$, $\tau^{VLD}$ from Eq. (44) is closely approximated by[39]

$$\tau_{as}^{VLD} \sim \frac{\tau_0 \pi e^\sigma}{4\alpha\sigma\sqrt{\delta(1+\delta)}}. \qquad (45)$$

Equation (45) follows from Eqs. (25), (58), and (81) for $h = 0$.

## VI. COMPARISON OF ANALYTICAL AND NUMERICAL RESULTS

We can now compare the reversal time $\tau^{VLD}$ for uniaxial and biaxial anisotropy from the exact integral solutions of Eqs. (9), (33), and (43), both with the asymptotic VLD escape rate $\tau_{as}^{VLD}$ and with the inverse of the smallest nonvanishing eigenvalue $\lambda_1$ of the Fokker-Planck operator, Eq. (3). The asymptotic escape rate is given in general by Eq. (35) and its particular cases embodied in Eqs. (53)-(55), (56)-(58), (69), and (81) below, while the eigenvalue is calculated numerically by the matrix continued fraction method.[13,33,39] All the calculations have been done for $\alpha = 0.001$ corresponding to the true VLD limit, $\alpha \Delta_{E_C} \ll 1$ for all values of the barriers which are used. For uniaxial and biaxial anisotropies, the reversal times are shown in Figs. 3 and 4, respectively, as functions of the inverse temperature parameter $\sigma$ for various values of the field parameter $h$ and typical values of the other model parameters. Clearly, for $\sigma > 5$, the *temperature dependence* of the reversal time has the customary Arrhenius behavior, i.e., $\tau^{\text{VLD}} \sim e^{\sigma(\varepsilon_C - \varepsilon_A^-)}$. This expression represents exponential increase with decreasing temperature; the slope of $\tau(T^{-1})$ being markedly dependent on $h$ because the barrier height of the shallow well is strongly influenced by this parameter as it significantly decreases with increasing $h$. In contrast for $\sigma < 3$, the behavior of $\tau^{\text{VLD}}(T^{-1})$ may deviate considerably from the Arrhenius behavior. Apparently, $\tau^{\text{VLD}}$ and $\lambda_1^{-1}$ lie very close to each other for virtually all $\sigma$. Furthermore in the high barrier limit, $\tau_{as}^{\text{VLD}}$ from the asymptotic Eq. (35) and its particular cases Eqs. (55), (69), (58), and (81) provide an accurate approximation to both $\lambda_1^{-1}$ and $\tau^{\text{VLD}}$. However, for $\sigma < 3$, $\tau_{as}^{\text{VLD}}$ deviates considerably from both of these so that it cannot be used to calculate the reversal time.

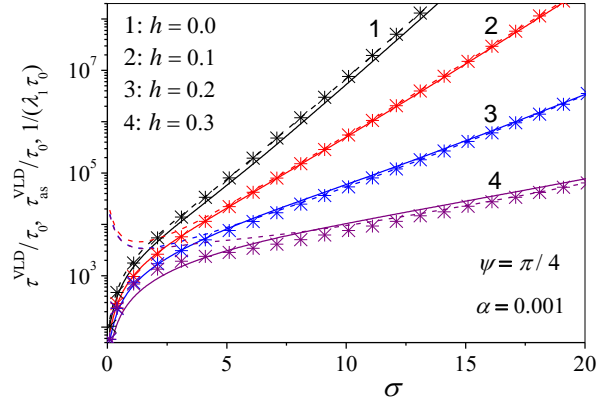

FIG. 3. Reversal time vs. the barrier height or inverse temperature parameter $\sigma = vK/(kT)$ for uniaxial anisotropy superimposed on the Zeeman term, Eq. (26), for various values of the field parameter $h$ with $\psi = \pi/4$. Solid line: $\tau^{\text{VLD}}$ from Eqs. (9), (33), and (65)-(68). Asterisks: numerical solution for the inverse of the smallest nonvanishing eigenvalue $1/(\tau_0 \lambda_1)$ of the Fokker-Planck operator in Eq. (3).[13,33] Dashed line: the VLD asymptotic Eqs. (6) (for $h = 0$ only) and general cases Eqs. (35), (53)-(55), (69).

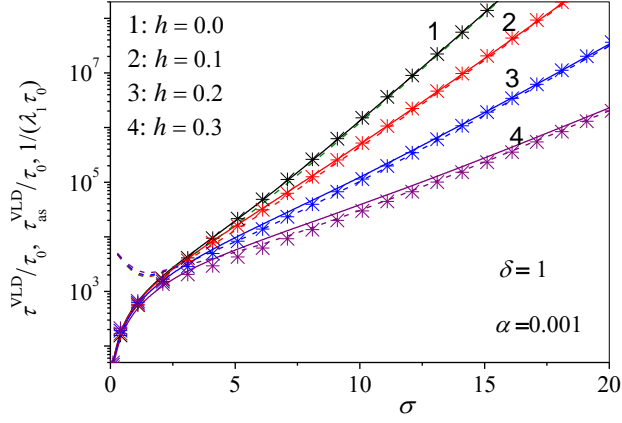

FIG. 4. Reversal time vs. the barrier height parameter $\sigma = vK/(kT)$ for biaxial anisotropy, Eq. (38), for various values of the field parameter $h$ with $\delta = 1$. Solid line: $\tau^{\text{VLD}}$ from Eqs. (9), (43), (78), and (80). Asterisks: numerical solution for $1/(\tau_0 \lambda_1)$.[13,39] Dashed line: the VLD asymptotic Eqs. (35), (56)-(58), and (81).

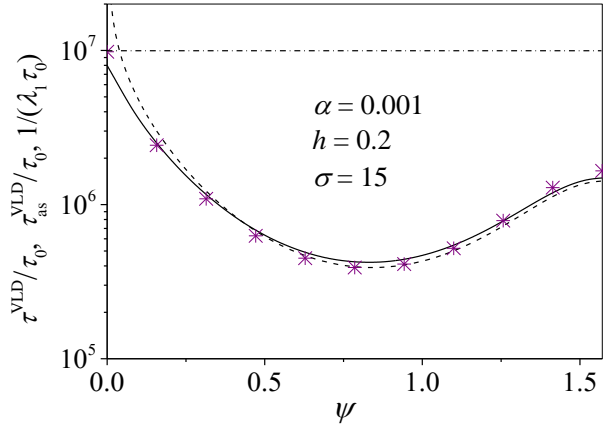

FIG. 5. Reversal time vs. oblique field angle $\psi$ for uniaxial anisotropy superimposed on the Zeeman term Eq. (26) and for $\sigma = 15$ and $h = 0.2$. Solid line: $\tau^{\text{VLD}}$ from Eqs. (9), (33), and (65)-(68). Asterisks: numerical solution for $1/(\tau_0 \lambda_1)$.[13,33] Dashed line: the VLD asymptotic Eqs. (35), (53)-(55), (69). Dashed-dotted line: Brown's Eq. (6).

To illustrate the problem of *uniaxial crossovers* encountered in calculating the reversal time $\tau^{\text{VLD}}$ via escape rate theory,[14,24] i.e., in the parameter range where departures from axial symmetry are small, we plot in Fig. 5 $\tau^{\text{VLD}}$ for uniaxial anisotropy superimposed on the Zeeman term [Eq. (26)] as a function of the oblique angle $\psi$ for $\sigma = 15$ and $h = 0.2$. In the interval $0 \leq \psi \leq \pi/2$ [note that $\tau(\psi) = \tau(\pi - \psi)$], $\tau^{\text{VLD}}$ strongly depends on $\psi$ having a minimum at $\psi \approx \pi/4$ and deviating considerably from the uniaxial asymptotic equation (6). However, in the limit of uniaxial crossover $\psi \to 0$, $\tau_{\text{as}}^{\text{VLD}}$ from the asymptotic equation (35) diverges because the dimensionless action $\Delta_{\varepsilon_C}^{\pm} \to 0$. Hence that equation cannot be used for the calculation of the reversal time. In contrast, the MFPT equations (9) and (33) yield values very close both to the numerical results and Brown's asymptote Eq. (6).

## VII. CONCLUSION

We have derived analytic formulas for the magnetization reversal time of nanomagnets in the VLD range. Our principal result is the general equation (20) yielding the reversal time via quadratures which can, in principle, be evaluated for *any* anisotropy. Yet another merit of Eq. (20) is that it is valid in parameter ranges, where escape rate equations such as Eqs. (6), (10), (35), and (45) do not apply at all, i.e., both for *low barriers* and *uniaxial crossovers*.[14,24] Furthermore, for certain anisotropies such as uniaxial and biaxial anisotropies, etc. the VLD reversal time can be given in terms of integrals of known functions, e.g., Eqs. (36), (37), and (44). Equation (20) is also valuable as a benchmark solution with which numerical calculations of the reversal time from the magnetic Langevin and/or Fokker-Planck equation in the VLD limit must agree. Finally, the method we have given can also be applied, with some modifications, to thermal agitation in magnetization dynamics in nanomagnets driven by spin-polarized currents.[12,28] Here a current of spin-polarized electrons is capable of applying nonconservative torques to the magnetization $\mathbf{M}$. In the spin-transfer-torque case, the stochastic dynamics of the magnetization $\mathbf{M}$ under VLD conditions are governed by an energy-controlled diffusion equation similar to Eq. (15) save for the spin-torque terms.[27,28] However, due to these terms the Gilbert damping may be overcome so that reversal of the magnetization becomes possible *in the absence* of thermal fluctuations. As far as spin-transfer torque and thermal fluctuations are concerned, the overall situation, albeit more complicated, is in some way reminiscent of that occurring in the resistively shunted junction (RSJ) model[13] of a Josephson junction which is an electric analog of the motion of a Brownian particle in a tilted periodic potential. Now just as the bias current in the junction, which constitutes a *nonconservative* electrical source giving rise to the motion in a tilted cosine periodic potential, ensures that the stationary distribution is no longer the Boltzmann distribution, in like manner the stationary distribution in a ferromagnet subjected to spin polarized current is no longer Boltzmann. Rather it depends both on the spin-polarized current and damping analogous to the dependence of the stationary distribution in the RSJ model on the bias current or tilt parameter. Some of the consequences include the switching time being systematically smaller than Brown's intrinsic thermally activated time in the low damping regime, and that the damping and external current parameter now governs the effective barrier heights so that the effect of the spin-polarized current may be as much as orders of magnitude.[41] The corresponding energy-controlled diffusion equations[12,28] can be analyzed by the methods outlined here.

We thank P.-M. Déjardin, L. Dowling and D. Jordan for helpful conversations. We would like to thank FP7-PEOPLE-Marie Curie Actions - International Research Staff Exchange Scheme (Project No. 295196 DMH) for financial support. W.T.C. also acknowledges the French Embassy in Ireland for enabling a research visit to Perpignan.

## APPENDIX A: CALCULATION OF $E_A$, $E_C$, AND $f_A$

In order to calculate the precession frequency in the well precession frequency $f_A$ and the well and saddle energies $E_A$ and $E_C$, it is supposed[8,13,14] that the free energy $E(\mathbf{M})$ of a nanoparticle has a multistable structure with minima at $\mathbf{n}_A^+$ and $\mathbf{n}_A^-$ separated by a potential barrier with a saddle point at $\mathbf{n}_C$ (see, e.g. Fig. 1). If $\mathbf{M}$ is close to the stationary points $\mathbf{n}_A^+$, $\mathbf{n}_A^-$, and $\mathbf{n}_C$ and $(u_1^{(p)}, u_2^{(p)}, u_3^{(p)})$, $p = A\pm, C$, denote the direction cosines of $\mathbf{M}$, then $E(\mathbf{M})$ can be approximated to second order in $u_1^{(p)}$ and $u_2^{(p)}$ via the Taylor series

$$E = E_p + \frac{1}{2}\left[c_1^{(p)}\left(u_1^{(p)}\right)^2 + c_2^{(p)}\left(u_2^{(p)}\right)^2\right] + \ldots. \tag{46}$$

To determine the expansion coefficients $c_1^{(p)}$, $c_2^{(p)}$, and $E_p$, we recall that the transformation matrix $\mathbf{R}^{(p)}$ relating the reference polar coordinate system $P$ and a new polar coordinate system $P'$ with the origin at the stationary point $\mathbf{n}_p$, is defined as[13,14]

$$\mathbf{R}^{(p)} = \begin{pmatrix} \cos\varphi_p \cos\vartheta_p & \sin\varphi_p \cos\vartheta_p & -\sin\vartheta_p \\ -\sin\varphi_p & \cos\varphi_p & 0 \\ \cos\varphi_p \sin\vartheta_p & \sin\varphi_p \sin\vartheta_p & \cos\vartheta_p \end{pmatrix},$$

so that the relationship between the direction cosines $u_n^{(p)}$ and $u_m'^{(p)}$ in the systems $P$ and $P'$ is given by

$$u_n^{(p)} = R_{1n}^{(p)} u_1'^{(p)} + R_{2n}^{(p)} u_2'^{(p)} + R_{3n}^{(p)} u_3'^{(p)} \tag{47}$$

($n = 1, 2, 3$). Because $u_3'^{(p)} \approx 1 - (u_1'^{(p)2} + u_2'^{(p)2})/2$, $E_p$ and $c_n^{(p)}$ ($n = 1, 2$) can be evaluated from Eqs. (46) and (47) as

$$E_p = E_p(u_1^{(p)}, u_2^{(p)})\Big|_{u_1'^{(p)}, u_2'^{(p)}=0}, \quad c_n^{(p)} = \frac{\partial^2 E}{\partial u_n'^{(p)2}}\Big|_{u_1'^{(p)}, u_2'^{(p)}=0} \tag{48}$$

The well precession frequencies $f_A^\pm$ are then defined via the Taylor expansion coefficients in Eq.(46) as

$$f_A^\pm = \frac{\gamma kT}{2\pi v \mu_0 M_S} \sqrt{c_1^{(A\pm)} c_2^{(A\pm)}}. \tag{49}$$

For uniaxial anisotropy in the presence of an uniform magnetic field $\mathbf{H}_0$ applied at an angle $\psi$ with respect to the easy axis, Eq. (26), the calculation yields[13,14]

$$E_p = -\sigma\left[\cos^2\vartheta_p + 2h\cos(\vartheta_p - \psi)\right], \tag{50}$$

$$c_1^{(p)} = 2\sigma\left[\cos 2\vartheta_p + h\cos(\vartheta_p - \psi)\right], \tag{51}$$

$$c_2^{(p)} = 2\sigma\left[\cos^2\vartheta_p + h\cos(\vartheta_p - \psi)\right], \tag{52}$$

where $\vartheta_p$ are the solutions of the trigonometric equation $\sin 2\vartheta = 2h\sin(\psi - \vartheta)$. The latter equation may be rewritten as the quartic equation[13,14,31]

$$(x + h\cos\psi)^2(1 - x^2) = (xh\sin\psi)^2$$

with $x = \cos\vartheta$. The roots of this quartic equation, viz.,

$$-1 \leq x_1 = \cos\vartheta_A^- < x_2 = \cos\vartheta_C^- < x_3 < x_4 = \cos\vartheta_A^+ \leq 1,$$

have a complicated algebraic form (see Ref. 30 for details). However, they can be written[34] as converging Taylor series to any desired order of $h$ allowing one to calculate $f_A$, $\varepsilon_A^\pm = E_A^\pm/\sigma$, and $\varepsilon_C = E_C/\sigma$ as

$$\varepsilon_A^\pm = -1 \mp 2h\cos\psi - h^2\sin^2\psi \pm \frac{h^3}{2}\sin 2\psi \sin\psi - \frac{h^4}{4}\sin^2 2\psi + \ldots, \tag{53}$$

$$\varepsilon_C = -2h\sin\psi + h^2\cos^2\psi + \frac{h^3}{2}\sin 2\psi \cos\psi + \frac{h^4}{4}\sin^2 2\psi + \ldots, \tag{54}$$

$$f_A^\pm = \frac{1}{2\pi\tau_0}\left[1 \pm h\cos\psi - \frac{h^2}{2}\sin^2\psi \pm \frac{3}{2}h^3\cos\psi\sin^2\psi - \frac{h^4}{16}(21 + 19\cos 2\psi)\sin^2\psi\right], \tag{55}$$

with $h < h_c(\psi) \leq 1$.

For biaxial anisotropy, Eq. (38), the calculation of $E_p$ and $c_n^{(p)}$ yields[13,14,39]

$$E_p = \sigma(\sin^2\vartheta_p - 2h\cos\vartheta_p), \quad c_1^{(p)} = 2\sigma(\cos 2\vartheta_p + h\cos\vartheta_p), \quad c_2^{(p)} = 2\sigma(\delta + \cos^2\vartheta_p + h\cos\vartheta_p),$$

where $\vartheta_p$ are the solutions of the equation $\partial_\vartheta V|_{\varphi=\pi/2} = 0$. These are $\vartheta_A^+ = 0$, $\vartheta_A^- = \pi$, and $\cos\vartheta_C = -h$. Thus

$$\varepsilon_A^\pm = E_A^\pm/\sigma = -1 \mp 2h, \tag{56}$$

$$\varepsilon_C = E_C/\sigma = h^2, \tag{57}$$

$$f_A^\pm = \frac{1}{2\pi\tau_0}\sqrt{(1 \pm h + \delta)(1 \pm h)}. \tag{58}$$

**APPENDIX B: CALCULATION OF $f_\varepsilon^\pm$ AND $S_\varepsilon^\pm$ FOR UNIAXIAL ANISOTROPY**

In order to evaluate the energy-dependent precession frequency $f_\varepsilon^+$ and the action $S_\varepsilon^+$ in the deeper well, we first introduce the parameter $p_\varepsilon$ defined as

$$(u_Z + h\cos\psi)^2 + 2h\sin\psi u_X = (h\cos\psi)^2 - \varepsilon = p_\varepsilon^2.$$

Next we introduce a new function $u(t)$ related to $u_X(t), u_Y(t)$, and $u_Z(t)$ via

$$u_X = \frac{p_\varepsilon^2}{2h\sin\psi}(1-u^2), \tag{59}$$

$$u_Y = \sqrt{1-(p_\varepsilon u - h\cos\psi)^2 - \frac{p_\varepsilon^4}{4h^2\sin^2\psi}(1-u^2)^2}, \tag{60}$$

$$u_Z = -h\cos\psi + p_\varepsilon u, \tag{61}$$

Then Eq. (29) becomes

$$\frac{du}{dt} = \frac{p_\varepsilon}{2\tau_0}\sqrt{(e_1-u)(u-e_2)(u-e_3)(u-e_4)}, \tag{62}$$

where $e_1$, $e_2$, $e_3$, and $e_4$ are the roots of the fourth-order polynomial $\Phi(u)$ given by

$$\Phi(u) = 4h^2 p_\varepsilon^{-4}\sin^2\psi\left[1-(p_\varepsilon u - h\cos\psi)^2\right] - (1-u^2)^2.$$

We do not give here the rather complicated explicit equations for $e_1$, $e_2$, $e_3$, and $e_4$ because using Mathematica® these roots can be easily calculated both analytically and numerically. Noting that in the deeper well, $u$ varies in the interval $e_1 \leq u \leq e_2$ and[43]

$$\int_u^{e_1}\frac{dx}{\sqrt{(e_1-x)(x-e_2)(x-e_3)(x-e_4)}} = \frac{2F(\varphi|m_\varepsilon)}{\sqrt{(e_1-e_3)(e_2-e_4)}}, \tag{63}$$

where $F(\varphi|m_\varepsilon)$ is the incomplete elliptic integral,[37]

$$\sin\varphi = \sqrt{\frac{(e_2-e_4)(e_1-u)}{(e_1-e_2)(u-e_4)}}, \quad m_\varepsilon = \frac{(e_1-e_2)(e_3-e_4)}{(e_1-e_3)(e_2-e_4)},$$

we have the explicit expression

$$u(t) = e_4\frac{b_\varepsilon + \text{sn}^2(\omega_\varepsilon t + w|m_\varepsilon)}{a_\varepsilon + \text{sn}^2(\omega_\varepsilon t + w|m_\varepsilon)}, \tag{64}$$

where $\text{sn}(u|m)$ is Jacobi's elliptic function[37] with real period $4K(m)$, $K(m)$ is the complete elliptic integral of the first kind,[37] $w$ is an initial phase, and

$$a_\varepsilon = \frac{e_2-e_4}{e_1-e_2}, \quad b_\varepsilon = \frac{e_1}{e_4}a_\varepsilon, \quad \omega_\varepsilon = \frac{p_\varepsilon}{4\tau_0}\sqrt{(e_1-e_3)(e_2-e_4)}.$$

Now, the energy-dependent precession frequency $f_\varepsilon^+$ of the magnetization is

$$f_\varepsilon^+ = \frac{p_\varepsilon\sqrt{(e_1-e_3)(e_2-e_4)}}{16\tau_0 K(m_\varepsilon)}, \tag{65}$$

while $S_\varepsilon^+$ from Eq. (34) is given by [noting Eqs. (27)-(29) and (59)-(61)]

$$S_\varepsilon^+ = \frac{\tau_0\sigma(4h^2\sin^2\psi - p^4)}{2f_\varepsilon^+} + \frac{\tau_0\sigma p^2}{2}\int_0^{1/f_\varepsilon^+}\left\{4ph\cos\psi\left[u(t)+u^3(t)\right]\right. \tag{66}$$
$$\left. + \left[4 - 4h^2\cos^2\psi - 2p^2\right]u^2(t) - p^2u^4(t)\right\}dt.$$

The calculation of $S_\varepsilon^+$ from Eq. (66) thus reduces to the evaluation of integrals of the form

$$f_\varepsilon^+ \int_0^{1/f_\varepsilon^+} \left[ \frac{b_\varepsilon + \text{sn}^2(\omega_\varepsilon t + w | m_\varepsilon)}{a_\varepsilon + \text{sn}^2(\omega_\varepsilon t + w | m_\varepsilon)} \right]^N dt = \sum_{n=0}^{N} \frac{N!(b_\varepsilon - a_\varepsilon)^n}{n!(N-n)!} I(n) \tag{67}$$

for $N = 1, 2, 3,$ and 4, where the integrals $I(n)$ are defined by

$$I(n) = f_\varepsilon^+ \int_0^{1/f_\varepsilon^+} \frac{dt}{\left[ a_\varepsilon + \text{sn}^2(\omega_\varepsilon t + w | m_\varepsilon) \right]^n}. \tag{68}$$

For $n = 1, 2, 3,$ and 4, the integrals $I(n)$ can be expressed in terms of the complete elliptic integrals as (using the table of integrals from Ref. 43)

$$I(1) = \frac{\Pi(-a_\varepsilon^{-1} | m_\varepsilon)}{a_\varepsilon K(m_\varepsilon)},$$

$$I(2) = \frac{1}{2a_\varepsilon(1+a_\varepsilon)} \left[ -1 + \frac{E(m_\varepsilon)}{(1+a_\varepsilon m_\varepsilon) K(m_\varepsilon)} + \frac{1 + 2a_\varepsilon + a_\varepsilon m_\varepsilon (3a_\varepsilon + 2)}{(a_\varepsilon + m_\varepsilon)} I(1) \right],$$

$$I(3) = \frac{3 + 9a_\varepsilon^2 m + 6a(1+m)}{4a_\varepsilon (1+a_\varepsilon)(1+a_\varepsilon m_\varepsilon)} I(2) - \frac{1 + m + 3am}{2a_\varepsilon(1+a_\varepsilon)(1+a_\varepsilon m_\varepsilon)} I(1) + \frac{m}{4a_\varepsilon(1+a_\varepsilon)(1+a_\varepsilon m_\varepsilon)},$$

$$I(4) = \frac{6 + 16a_\varepsilon^2 m + 11a(1+m)}{6a_\varepsilon(1+a_\varepsilon)(1+a_\varepsilon m_\varepsilon)} I(3) + \frac{3 + 57a_\varepsilon^2 m + 22a(1+m)}{24a_\varepsilon^2(1+a_\varepsilon)(1+a_\varepsilon m_\varepsilon)} I(2) + \frac{2(1+m+9am)I(1) - m}{24a_\varepsilon^2(1+a_\varepsilon)(1+a_\varepsilon m_\varepsilon)},$$

where $E(m)$ and $\Pi(n|m)$ are the complete elliptic integrals of the second and third kind, respectively.[37]

In order to calculate the reversal time $\tau^{\text{VLD}}$, we actually need only $f_\varepsilon^+$ and $S_\varepsilon^+$, i.e., Eqs. (65) and (66), corresponding to the deeper well, because $S_\varepsilon^-$ and $f_\varepsilon^-$ for the shallow well can be formally obtained by replacing *in all the equations* $h$ by $-h$. Hence in Eq. (9), we have $\tau_+ = \tau_+(h)$ and $\tau_- = \tau_+(-h)$. We remark that at the saddle point $\varepsilon_C$, $S_{\varepsilon_C}^\pm$ can be evaluated from Eq. (23) via converging Taylor series to any desired order of $h$[34]

$$S_{\varepsilon_C}^\pm = \sigma \sqrt{|h| \sin \psi} \left[ 16 - \frac{104}{3} |h| \sin \psi + h^2 (1 - 21 \cos 2\psi) + \ldots \right] \pm 2\pi \sigma h^2 \sin 2\psi (4 - 3|h| \sin \psi + \ldots). \tag{69}$$

The above results can be simplified for axial symmetry, $\psi \to 0$, where the frequency $f_\varepsilon^+ \big|_{\psi \to 0}$ from Eq. (65) and the actions $S_\varepsilon^\pm \big|_{\psi \to 0}$ from Eqs. (66) reduce to

$$f_\varepsilon^+ \big|_{\psi \to 0} = \frac{1}{2\pi \tau_0} \sqrt{h^2 - \varepsilon}. \tag{70}$$

$$S_\varepsilon^\pm \big|_{\psi \to 0} = 4\pi \sigma \sqrt{h^2 - \varepsilon} \left( \varepsilon - 2h^2 + 1 \pm 2h \sqrt{h^2 - \varepsilon} \right). \tag{71}$$

# APPENDIX C: CALCULATION OF $f_\varepsilon^\pm$ AND $S_\varepsilon^\pm$ FOR BIAXIAL ANISOTROPY

Noting that Eqs. (42) and (31) lead to the equality

$$\left(u_Z + \frac{h}{\delta+1}\right)^2 + \frac{\delta}{\delta+1}u_Y^2 = \frac{\delta-\varepsilon}{\delta+1} + \frac{h^2}{(\delta+1)^2} = p_\varepsilon^2,$$

we can again introduce a new function $u(t)$ related to $u_X(t), u_Y(t)$, and $u_Z(t)$ via

$$u_Z(t) = p_\varepsilon u(t) - h(\delta+1)^{-1}, \tag{72}$$

$$u_Y(t) = p_\varepsilon \sqrt{\left(1+\delta^{-1}\right)\left[1-u^2(t)\right]}, \tag{73}$$

$$u_X(t) = p_\varepsilon \sqrt{\delta^{-1}[u(t)-e_+][u(t)-e_-]}, \tag{74}$$

where

$$e_\pm = -\frac{h\delta}{p_\varepsilon(\delta+1)} \pm \frac{\sqrt{h^2-\varepsilon}}{p_\varepsilon}. \tag{75}$$

In the deeper well, $u$ varies in the interval $e_+ \leq u \leq 1$, while in the shallow well it varies in the interval $-1 \leq u \leq e_-$. By substituting Eqs. (72)-(74) into Eq. (41), we have

$$\frac{du}{dt} = -\frac{p_\varepsilon\sqrt{\delta+1}}{\tau_0}\sqrt{(1-u^2)(u-e_+)(u-e_-)}. \tag{76}$$

The solution of Eq. (76) is given in terms of Jacobi's elliptic function $\mathrm{sn}(u|m),^{37}$ viz.,

$$u(t) = \frac{a_\varepsilon - \mathrm{sn}^2(\omega_\varepsilon t + w|m_\varepsilon)}{a_\varepsilon + \mathrm{sn}^2(\omega_\varepsilon t + w|m_\varepsilon)}. \tag{77}$$

where,

$$a_\varepsilon = \frac{1+e_+}{1-e_+}, \quad m_\varepsilon = \frac{(1+e_-)(1-e_+)}{(1+e_+)(1-e_-)},$$

$$\omega_\varepsilon = \frac{p_\varepsilon}{2\tau_0}\sqrt{(\delta+1)(1+e_+)(1-e_-)}.$$

Note that $0 \leq m_\varepsilon \leq 1$ for $-1+2h < \varepsilon < h^2$ and $-1 < m_\varepsilon \leq 0$ for $-1-2h < \varepsilon < -1+2h$.

Now, the precession frequency $f_\varepsilon^+$ is

$$f_\varepsilon^+ = \frac{\omega_\varepsilon}{4K(m_\varepsilon)} = \frac{p_\varepsilon\sqrt{(\delta+1)(1+e_+)(1-e_-)}}{8\tau_0 K(m_\varepsilon)}, \tag{78}$$

while $S_\varepsilon^+$ is given by [noting Eqs. (39)-(41) and (72)-(74)]

$$S_\varepsilon^+ = \frac{2\sigma p_\varepsilon^2}{\tau_0}\int_0^{1/f_\varepsilon^+}\left\{(1+\delta)\varepsilon - h^2 + 2hp_\varepsilon(1+\delta)u(t) + \left(1+\delta-h^2\right)u^2(t)\right\}dt. \tag{79}$$

Next with Eqs. (67), we have

$$S_\varepsilon^+ = \frac{16\sigma p_\varepsilon K(m_\varepsilon)\sqrt{1+\delta}}{\sqrt{(1+e_+)(1-e_-)}}\left\{\varepsilon - \frac{h^2}{1+\delta} - 2hp_\varepsilon\left[1 - 2\frac{\Pi(-a_\varepsilon^{-1}|m_\varepsilon)}{K(m_\varepsilon)}\right]\right.$$
$$\left. + \frac{1+\delta-h^2}{1+\delta}\left[\frac{1-a_\varepsilon}{1+a_\varepsilon} + 2\frac{a_\varepsilon E(m_\varepsilon)-(1-a_\varepsilon^2 m_\varepsilon)\Pi(-a_\varepsilon^{-1}|m_\varepsilon)}{(1+a_\varepsilon)(1+a_\varepsilon m_\varepsilon)K(m_\varepsilon)}\right]\right\}$$
(80)

Again, for the shallow well, $S_\varepsilon^-$ and $f_\varepsilon^-$ can be formally obtained simply by replacing in all the equations $h$ by $-h$. Thus, we have $\tau_+(h)$ and $\tau_- = \tau_+(-h)$. At the saddle point $\varepsilon_C$, $S_{\varepsilon_C}^\pm$ can be evaluated as

$$S_{\varepsilon_C}^\pm = 8\delta\sigma\left(1 - \frac{h^2}{1+\delta}\right)\left\{\sqrt{\frac{1-h^2}{\delta}} + \frac{h}{\sqrt{1+\delta}}\arctan\left[\frac{h}{\sqrt{(1-h^2)(1+\delta^{-1})}}\right] \pm \frac{h\pi}{2}\right\}. \quad (81)$$

The above results can be simplified for two particular cases of interest, namely, for axial symmetry, $\delta \to 0$, and for zero external field, $h \to 0$. For $\delta \to 0$, $f_\varepsilon^+\big|_{\delta \to 0}$ from Eq. (78) and $S_\varepsilon^+\big|_{\delta \to 0}$ from Eq. (79) are given by Eqs. (70) and (71), respectively. For $h \to 0$, we have from Eqs. (78) and (80)

$$f_\varepsilon^+\big|_{h\to 0} = \frac{\sqrt{\delta(1+\varepsilon)}}{8\tau_0 K(a_\varepsilon^{-2})} = \frac{\sqrt{\delta-\varepsilon}}{4\tau_0 K\left(\frac{\delta+\delta\varepsilon}{\delta-\varepsilon}\right)}, \quad (82)$$

$$S_\varepsilon^+\big|_{h\to 0} = S_\varepsilon^-\big|_{h\to 0} = S_\varepsilon = \frac{16\sigma\sqrt{\delta-\varepsilon}}{1+e_+}\left\{\left(\varepsilon - \frac{a_\varepsilon-1}{a_\varepsilon+1}\right)K(a_\varepsilon^{-2}) + \frac{2a_\varepsilon^2 E(a_\varepsilon^{-2})}{(1+a_\varepsilon)^2}\right\}$$
$$= 8\sigma\sqrt{\delta-\varepsilon}\left[E\left(\frac{\delta+\delta\varepsilon}{\delta-\varepsilon}\right) + \varepsilon K\left(\frac{\delta+\delta\varepsilon}{\delta-\varepsilon}\right)\right]. \quad (83)$$

Here we have used known equations from the theory of elliptic functions, namely,[42]

$$K(a_\varepsilon^{-2}) = \frac{1}{2}\left(1+\sqrt{m'}\right)K(1-m')$$

and

$$E(a_\varepsilon^{-2}) = \frac{1}{1+\sqrt{m'}}\left[\sqrt{m'}K(1-m') + E(1-m')\right],$$

where $\sqrt{m'} = (a_\varepsilon-1)/(a_\varepsilon+1)$.

```mathematica
ep[δ_, h_, ϵ_] := (-(h*δ)/(1+δ) + √(h^2 - ϵ))/√(δ/(1+δ) - ϵ/(1+δ) + h^2/(1+δ)^2); ap[δ_, h_, ϵ_] := (1 - ep[δ, h, ϵ])/(1 + ep[δ, h, ϵ]); me[δ_, h_, ϵ_] := (1 - ep[δ, -h, ϵ])/(1 + ep[δ, h, ϵ]) * (1 - ep[δ, h, ϵ])/(1 + ep[δ, -h, ϵ]);

dep[δ_, h_, α_, ϵ_] := (16*α*EllipticK[me[δ, h, ϵ]] * √(δ - ϵ + h^2/(1+δ)))/√((1 + ep[δ, h, ϵ])*(1 + ep[δ, -h, ϵ])) *
  ((ϵ - h^2/(δ+1) - 2*h*√(δ/(1+δ) - ϵ/(1+δ) + h^2/(1+δ)^2)) * (1 - 2*EllipticPi[-ap[δ, h, ϵ], me[δ, h, ϵ]]/EllipticK[me[δ, h, ϵ]]) +
    (1 - h^2/(1+δ))/(ap[δ, h, ϵ] + 1) * (ap[δ, h, ϵ] - 1 + 2 * (ap[δ, h, ϵ]*EllipticE[me[δ, h, ϵ]] - (ap[δ, h, ϵ]^2 - me[δ, h, ϵ])*EllipticPi[-ap[δ, h, ϵ], me[δ, h, ϵ]])/((ap[δ, h, ϵ] + me[δ, h, ϵ])*EllipticK[me[δ, h, ϵ]])));

tp[σ_, δ_, h_, α_] := 8*σ*NIntegrate[Exp[σ*ee]/dep[δ, h, α, ee] * NIntegrate[(EllipticK[me[δ, h, z]]*Exp[-σ*z])/√((δ - z + h^2/(1+δ))*(1 + ep[δ, h, z])*(1 + ep[δ, -h, z])), {z, -1 - 2*h, ee}], {ee, -1 - 2*h, h^2}];

tm[σ_, δ_, h_, α_] := 8*σ*NIntegrate[Exp[σ*ee]/dep[δ, -h, α, ee] * NIntegrate[(EllipticK[me[δ, -h, y]]*Exp[-σ*y])/√((δ - y + h^2/(1+δ))*(1 + ep[δ, h, y])*(1 + ep[δ, -h, y])), {y, -1 + 2*h, ee}],
  {ee, -1 - 2*h, h^2}];

tau[σ_, δ_, h_, α_] := Block[{qm, qp}, qp = tp[σ, δ, h, α]; qm = tm[σ, δ, h, α]; Return[(2*qm*qp)/(qm + qp)];];

TTT[σ_, α_] := (√Pi * σ)/α * NIntegrate[(e^(-z^2) * (Erfi[√σ] - Erfi[z]))/(σ - z^2), {z, 0, √σ}];

δ = 0.0000000000001; α = 0.01; σ = 21.; h = 0.;
sss = (2*σ)/α * NIntegrate[Exp[-σ*(z^2 + 2*h*z)]/(1 - z^2) * Exp[σ*(x^2 + 2*h*x)], {z, -h, 1}, {x, z, 1}]
dat = TTT[σ, α]
dat1 = tau[σ, δ, h, α]
```

Out[176]= 2.68354×10^10

Out[177]= 2.68354×10^10

Out[178]= 2.68354×10^10

```mathematica
ttp[σ_, α_] := (√(π*σ))/(2*α) * NIntegrate[(e^(σ*(ee-h^2)) * (Erfi[(1 + h)√σ] - Erfi[√(σ(h^2 - ee))]))/(√(h^2 - ee) * (ee + 1 - 2*h^2 + 2*h*√(h^2 - ee))), {ee, -1 - 2*h, h^2}];

ttm[σ_, α_] := (√(π*σ))/(2*α) * NIntegrate[(e^(σ*(ee-h^2)) * (Erfi[(1 - h)√σ] - Erfi[√(σ(h^2 - ee))]))/(√(h^2 - ee) * (ee + 1 - 2*h^2 - 2*h*√(h^2 - ee))), {ee, -1 + 2*h, h^2}];

ttau[σ_, α_] := (2*ttp[σ, α]*ttm[σ, α])/(ttp[σ, α] + ttm[σ, α]);
ti = Date[]
α = 0.01; h = 0.1; σ = 100.;
dat = ttp[σ, α]
qqq = (2*σ)/α * NIntegrate[Exp[-σ*(z^2 + 2*h*z)]/(1 - z^2) * Exp[σ*(x^2 + 2*h*x)], {z, -h, 1}, {x, z, 1}]
qp = (√(π*σ))/α * NIntegrate[Exp[-σ*(h + z)^2]/(1 - z^2) * (Erfi[(1 + h)√σ] - Erfi[(h + z)√σ]), {z, -h, 1}]

qm = (√(π*σ))/α * NIntegrate[Exp[-σ*(h + z)^2]/(1 - z^2) * (Erfi[(1 - h)√σ] + Erfi[(h + z)√σ]), {z, -1, -h}]
dat = ttm[σ, α]
ta = (2*qp*qm)/(qp + qm)
dat = ttau[σ, α]
as = √(Pi/σ)/(α*(1 - h^2) * ((1 + h)*Exp[-σ*(1 + h)^2] + (1 - h)*Exp[-σ*(1 - h)^2]))

tee = Date[] - ti
```

Out[58]= {2014, 4, 4, 10, 38, 19.1711936}

Out[60]= 2.87885×10^53

Out[61]= 2.87885×10^53

Out[62]= 2.87885×10^53

Out[63]= 1.53305×10^96

Out[64]= 1.53305×10^96

Out[65]= 3.06611×10^96

Out[66]= 3.06611×10^96

Out[67]= 2.99606×10^96

Out[58]= {0, 0, 0, 0, 0, 0.4260244}